\documentclass[12pt]{article}
\usepackage{amsmath}
\usepackage{graphicx,psfrag,epsf}
\usepackage{enumerate}
\usepackage{natbib}
\usepackage{url} 
\usepackage[hidelinks]{hyperref}

\newcommand{\blind}{0}

\begin{document}

\def\spacingset#1{\renewcommand{\baselinestretch}%
{#1}\small\normalsize} \spacingset{1}


\if0\blind
{
  \title{\bf The Digital-Energy Recomposition of the Kitchin Cycle:
Exploratory Evidence from U.S.\ IT and Irish Data-Centre Series (1995-2025)}
  \author{Bartosz Zi\'o{\l}ko\thanks{Corresponding author: \protect\url{https://home.agh.edu.pl/~bziolko/en.html}. The research presented in this paper received partial funding from the Polish Ministry of Science and Higher Education assigned to AGH University of Krakow. The auxiliary LLM/earnings-call citation \citet{ZiolkoPPRAI26} is an own forthcoming work by the same authors and is included for context only; the headline results of the present paper do not rely on it.},
    Adam Niedzia{\l}kowski 
    and
    Bartosz Szyma \\
    Faculty of Computer Science, AGH University of Krakow, Poland\\}
  \maketitle
 }\fi

\if1\blind
{
  \bigskip
  \bigskip
  \bigskip
  \begin{center}
    {\LARGE\bf The Digital-Energy Recomposition of the Kitchin Cycle:
Exploratory Evidence from U.S.\ IT and Irish Data-Centre Series (1995-2025)}
\end{center}
  \medskip
} \fi

\bigskip
\begin{abstract}
Using quarterly FRED and CSO Ireland data over 1995Q1-2025Q4, we examine whether U.S.\ IT investment and Irish data-centre electricity are recomposing the Kitchin cycle around energy- and compute-backed assets. Three findings stand out. First, a Quandt-Andrews sup-Wald test locates a significant break in the U.S.\ IT-GDP correlation at 2014Q3 (sup-W $= 9.69$; $r$ falls from $+0.45$ to $-0.12$); the Fisher $r$-to-$z$ test survives Bonferroni adjustment ($z = +3.11$, $p_{\text{Bonf}} = 0.006$), whereas the analogous test on the imposed Cloud-vs-AI boundary does not ($p_{\text{Bonf}} = 0.164$), so ``Strategic Decoupling'' pre-dates the AI Era by approximately six years. Second, in Ireland the synchronous AI-Era correlation between metered data-centre electricity and Real GDP is large ($r = +0.41$, $N = 16$) but \emph{collapses to $r \approx 0$ when Real GDP is replaced by Real GNP}, suggesting the headline is largely an artefact of the shared multinational tax-structure footprint between hyperscaler operations and Irish gross output. Third, the Irish cross-correlation profile is asymmetric in the opposite direction to a leading-indicator reading: the largest positive coefficients cluster on the $k < 0$ side (GDP precedes data-centre load), so compute capacity is better read as coincident or moderately lagging. Given the small AI-Era samples and the absence of a quarterly multinational-adjusted aggregate such as GNI*, all coefficients are exploratory; subject to these caveats, the U.S.\ Kitchin cycle is best dated to 2014Q3, and the Irish Energy-Data Nexus is dominated by multinational firm activity.
\end{abstract}

\noindent%
\vfill

\section{Introduction}
\label{sec:intro}

The study of economic periodicity has long been a cornerstone of macroeconomic forecasting and investment strategy. Among the established wave theories, the \textbf{Kitchin cycle}, first identified by Joseph Kitchin in 1923, remains one of the most influential frameworks for understanding short/medium-term economic fluctuations. In his seminal work, \textit{``Cycles and Trends in Economic Factors''}, Kitchin observed a minor cycle of approximately 40 months (3-5 years), driven by fluctuations in commercial inventories and capital investment.

For nearly a century, the Kitchin cycle served as a reference point for asset allocation and production planning. It operated on a simple feedback loop: a surge in demand led to an accumulation of physical inventory (e.g., steel, coal, finished goods), eventually resulting in market saturation, price corrections, and subsequent inventory liquidation. This ``inventory lag'' provided a predictable rhythm that investors used to time market entries and exits.

However, the structural realities of the 21st-century global economy differ from the industrial landscape of the 1920s. We argue that the traditional Kitchin cycle, while theoretically elegant, may be losing relevance in its original form. The transition from a manufacturing-based economy to a digital-first infrastructure plausibly shifts the relevant ``inventory'' from physical warehouses to digital ones, specifically to \textbf{compute power, semiconductor stockpiles, and energy contracts}.

This paper introduces a revised framework: the \textbf{``New Energy and Data Cycle.''} We document an empirical ``Strategic Decoupling'' of U.S.\ IT investment from GDP, and ask when, on the data, it begins. While our nominal AI Era window is 2021Q1-2025Q4, a Quandt-Andrews sup-Wald test on the U.S.\ IT-GDP correlation locates the structural break in 2014Q3, several years before the current wave of AI investment; the corresponding Fisher $r$-to-$z$ test on the pre/post-2014Q3 sub-samples survives Bonferroni adjustment ($z = +3.11$, $p_{\text{Bonf}} = 0.006$), in contrast to the analogous test on the imposed Cloud-vs-AI era boundary ($p_{\text{Bonf}} = 0.164$). The pattern we report is therefore best read as exploratory evidence consistent with a longer-running decoupling of IT capital formation from GDP, of which the AI Era is the latest segment, rather than as a phenomenon uniquely attributable to AI. In parallel, we examine whether energy consumption in data centers and semiconductor manufacturing behaves as a coincident or leading correlate of growth in the AI Era window.

Using quarterly data from the Republic of Ireland, a country with a globally significant share of data-center capacity, we examine whether the modern Kitchin cycle is better explained by the \textit{availability and consumption of electricity} than by physical goods. We acknowledge that Irish GDP statistics are heavily influenced by multinational firm activity \citep{fitzgerald2018}, a well-documented limitation revisited in Section~\ref{sec:limitations}; accordingly, the Irish evidence is presented as an illustrative, not a confirmatory, case study. Our results suggest that power purchase agreements (PPAs) and data center load deserve consideration alongside traditional stock-to-sales ratios as candidate barometers of economic health in data-intensive economies.

\section{Literature Review}

The study of economic periodicity has long been anchored in the framework established by \citet{kitchin1923}, who identified a ``minor'' cycle of approximately 40 months (3-5 years) driven by fluctuations in commercial inventories and capital investment. \citet{schumpeter1939} subsequently embedded the Kitchin cycle into a broader synthesis of short, medium (Juglar) and long (Kondratiev) waves, and \citet{burns1946} provided the methodological foundation for measuring such cycles empirically via reference-cycle dating and the inventory-to-sales ratio. Traditionally, this cycle operated on a feedback loop where demand surges led to the accumulation of physical stocks, such as steel and coal, eventually resulting in market saturation and liquidation. However, as the global economy transitions toward a digital-first infrastructure, the ``inventory'' at the heart of this cycle plausibly shifts from physical warehouses to digital ones, encompassing compute power and energy contracts.

The traditional Kitchin framework treats energy as a near-perfectly elastic input. The ``Energy-GDP Nexus'' literature suggests, by contrast, that in high-tech economies, energy constraints can dictate the upper bounds of the cycle \citep{stern2011}. As data centers evolve into a key element of digital industrial capacity, electricity consumption may shift from a coincident toward a leading role. The empirical literature on data-centre energy use is dominated by \citet{koomey2008}, which documented the doubling of global data-centre electricity demand between 2000 and 2005, and \citet{masanet2020}, which recalibrated the global data-centre energy estimate to about 200 TWh per year despite an order-of-magnitude increase in compute, by isolating efficiency gains in hyperscaler facilities. The \citet{iea2024} \textit{Electricity 2024} report extends this trajectory through 2026, projecting that AI-driven workloads, cryptocurrency, and data-centre cooling will roughly double global data-centre electricity demand within two years. These works frame our use of metered data-centre consumption as a candidate macroeconomic indicator: the question is whether the country-level cycle picks up the structural growth that these aggregate studies document.

This potential structural break can be understood through the lens of technological revolutions. \citet{perez2002} describes how financial capital and technological shifts create structural changes in economic rhythms, a pattern arguably mirrored in the current transition into the AI era. Consequently, the nature of capital itself appears to have changed: as \citet{haskel2017} argue, the rise of the intangible economy has decoupled growth from physical assets, making traditional inventory metrics less informative.

A critical component of this dematerialized inventory is the semiconductor sector. Unlike the 1920s steel inventories, the ``Silicon Cycle'' is characterized by extreme capital intensity and long lead times. This inherent lag is one possible explanation for the ``Strategic Decoupling'' observed in recent data; semiconductor manufacturing often follows a 4-to-5 year investment rhythm (Juglar-like) that overlaps and occasionally disrupts the shorter Kitchin cycle. The negative correlation observed between 2021 and 2025 may thus represent a ``pre-emptive investment phase,'' in which firms accumulate compute capacity largely independently of current GDP fluctuations to hedge against future supply shocks.

The shift toward AI-driven utility also calls for a re-evaluation of investment behavior through the lens of \textit{Real Options Theory} \citep{dixit1994}. From this perspective, IT investment in the AI era is not a reaction to current output, but a strategic ``option'' on future productivity. This is consistent with the observation that, despite macroeconomic headwinds, investment in GPU clusters and energy-backed infrastructure has remained resilient. The decoupling of growth from physical assets emphasised by \citet{haskel2017} is thus arguably extended by the virtualization of the inventory itself: from stockpiles of goods to stockpiles of ``potential computation.''

The challenge of measuring such cycles in real time is addressed in the methodology of \citet{aruoba2009}, whose work emphasises the use of high-frequency data to monitor business conditions, providing a statistical precedent for using non-traditional indicators like electricity consumption; recent work on large language models in financial analysis \citep{ZiolkoPPRAI26} is auxiliary to the energy-centred argument advanced here.

Taken together, the literature motivates the hypothesis that the Kitchin cycle has not disappeared but is being recomposed: traditional IT investment is increasingly decoupled from contemporaneous GDP, while semiconductor stockpiling and data-center capacity behave more like strategic, anti-cyclical assets. The analysis that follows is intended to probe, rather than confirm, this hypothesis.

\section{Methodology}
\label{sec:meth}
The core objective of this study is to examine the structural shift in the Kitchin cycle, moving from a traditional inventory-based model to a digital-energy framework. We employ a multi-stage quantitative approach to analyze the relationship between macroeconomic growth and technological infrastructure. The \emph{ex ante} ``Digital-Energy Kitchin Cycle'' framework underlying the empirical strategy below is reported as Figure~S1 in the Supplementary Materials; the analysis that follows tests, rather than assumes, that framework.

\subsection{Analytical Framework}
We segment the historical data into distinct technological eras (PC Era, Dot-com Boom, Web 2.0, Cloud Expansion, and the AI Era beginning 2021Q1) to examine whether correlation patterns differ across regimes. The era boundaries are taken from the descriptive technology-history literature and are imposed rather than estimated from the data; we acknowledge in Section~\ref{sec:limitations} that this is an exploratory choice and that formal structural-break tests would be a natural confirmatory extension. The primary focus is on the ``New Energy Hypothesis,'' which posits that in the current AI era, traditional IT investment indicators decouple from GDP, while energy-related metrics (specifically data center power consumption) emerge as candidate leading indicators.

\subsection{Statistical Techniques}
The following statistical methods are applied:
\begin{itemize}
    \item \textbf{Rolling Correlations:} a 20-quarter (5-year) window is used to track the dynamic evolution of IT-GDP synchronicity. The window length matches the standard 5-year Kitchin horizon but interacts uncomfortably with our short AI Era: the AI Era contains 20 quarterly observations, so within this window the rolling correlation collapses to a single end-point value once the window slides into the era, carrying roughly 3-4 ``effectively new'' end-points after 2021Q1, which we flag explicitly to discourage reading the trend as a continuous, well-identified estimate over the AI Era.
    \item \textbf{Cross-Correlation Analysis (Lead-Lag):} we test for lead-lag relationships by shifting variables between $-4$ and $+4$ quarters, in order to characterise whether technological variables behave as leading, lagging, or coincident indicators.
    \item \textbf{Log-Return Transformation:} all time-series are converted to quarterly log-returns ($\ln(x_t/x_{t-1})$) to focus on growth rates and reduce the influence of low-frequency trends. We run Augmented Dickey-Fuller (ADF) and KPSS tests on every series in both log-levels and log-returns (full results in \verb|stationarity_tests.csv| and the Supplementary Materials Section~S5). All eight log-level series are non-stationary; after log-differencing, ADF rejects the unit-root null at 5\% for every series, but the U.S.\ semiconductor production and import indices retain a KPSS rejection at $p = 0.01$ even on log-returns. We flag both as potential residual unit roots and discuss the implication for headline correlations involving semiconductor variables in Section~\ref{sec:limitations}.
    \item \textbf{Multiple Testing Correction:} we compute Pearson correlations across nine cross-correlation lags, six counterpart variables (of which three -- US IT investment, US electricity PPI, US semiconductor production -- are reported as headline series in Table~\ref{tab:correlations}; the remaining three, US semiconductor imports, US renewables-generation proxy, and Irish data-center metered electricity, enter the cross-lag, Granger and spectral panels), and multiple era windows, yielding $\sim 120$ tests in total. To control the false discovery rate at $q = 0.05$ we apply Benjamini-Hochberg adjustment across the pooled $p$-values. Tables~\ref{tab:correlations} and~\ref{tab:lags} therefore report both raw and FDR-adjusted significance.
    \item \textbf{Granger Causality:} where sample sizes allow, we perform \citet{granger1969} causality tests to evaluate whether electricity growth predicts future IT investment and GDP fluctuations. Given the small effective sample for the post-2021 window ($N=20$), Granger results are reported as descriptive rather than as definitive evidence of causation.
    \item \textbf{Frequency-Domain Analysis (Welch PSD):} to test whether the canonical 3-5 year Kitchin periodicity is present in the headline log-return series, we compute the Welch power spectral density via \texttt{scipy.signal.welch} with sampling frequency $f_s = 4$ cycles per year (quarterly data), a Hann window, segment length $n_{\text{perseg}} = \min(32, N)$, 50\% segment overlap, and constant detrending. The segment length is chosen short enough that even the Irish data-center series ($N = 39$ log-return observations) yields multiple non-overlapping segments, at the cost of a coarser frequency grid relative to the U.S.\ series ($N \approx 123$). We then report the period (in years) of the top three non-zero PSD peaks for each series in Section~\ref{sec:results-spectral}; the resulting peaks are interpreted against the canonical Kitchin band of 3-5 years (frequencies $0.20$-$0.33$ cycles per year). PSD estimates are point estimates only; we do not attach formal confidence bands, since the residual autocorrelation in quarterly log-returns and the short sample make HAC-corrected spectral inference an explicit caveat (Section~\ref{sec:limitations}).
\end{itemize}

\subsection{Data Acquisition and Preprocessing}
The empirical analysis utilizes quarterly observations retrieved from Federal Reserve Economic Data (FRED) and the Central Statistics Office (CSO) of Ireland. The sample window for the headline U.S.\ tables is frozen at 1995Q1-2025Q4, yielding $N = 122$ paired quarters after log-differencing; Irish series begin later (2015Q1 for CSO data-centre metered electricity). The pipeline processes U.S.\ real GDP (FRED: GDPC1), U.S.\ IT investment (B935RC1Q027SBEA), U.S.\ semiconductor production and import indices (CAPG3344S, IR21320), and Irish data-centre metered electricity (CSO: MEC02); full series details, source URLs and retrieval vintages are in the Supplementary Materials Section~S5. Monthly indicators are resampled to quarterly frequency using a mean-within-quarter approach.

\subsection{Statistical Transformations}
To focus on growth rates and ensure stationarity across long-term cycles, all time-series data undergo a log-return transformation:
\begin{equation}
    r_t = \ln(x_t / x_{t-1})
\end{equation}
This method aligns with established macroeconomic forecasting standards for identifying cyclical volatility rather than absolute level shifts.

\subsection{Temporal Segmentation and Rolling Analysis}
We segment historical data into six technological eras chosen to capture structural breaks in correlation patterns. The era boundaries are imposed rather than estimated, and we discuss the limits of this approach in Section~\ref{sec:limitations}.
\begin{enumerate}
    \item \textit{Pre-Tech Era} (1959Q1-1982Q4, $N=95$)
    \item \textit{PC Era} (1983Q1-1991Q1, $N=33$)
    \item \textit{Dot-com Boom} (1991Q2-2001Q1, $N=40$)
    \item \textit{Web 2.0 Era} (2001Q4-2007Q4, $N=24$)
    \item \textit{Cloud Expansion} (2009Q2-2019Q4, $N=42$)
    \item \textit{AI Era} (2021Q1-2025Q4, $N=20$ for U.S.\ series; $N \le 16$ for Irish data-center series, which ends 2024Q4)
\end{enumerate}
The Pre-Tech and PC Era windows are listed above for historical reference only and are \emph{not} reported as headline coefficients in this paper: the U.S.\ private fixed investment in information processing equipment series (FRED B935RC1Q027SBEA), which anchors Tables~\ref{tab:correlations} and~\ref{tab:decoupling}, begins in 1995Q1, so the headline era-by-era windows we report start at Dot-com Boom (1991Q2-2001Q1) and the effective IT-GDP correlation in that era is computed on $N=23$ paired log-returns. The two earlier eras enter only the historical-context narrative.

A rolling correlation window of 20 quarters (5 years) is retained to monitor the dynamic ``Strategic Decoupling'' of IT from GDP growth.

The era $N$ reported above refers to the maximum number of quarterly log-return observations available within each window (anchored to U.S.\ GDP, which is published since 1947Q1). The \emph{effective} $N$ used in any specific era-by-era correlation can be smaller when the counterpart series begins later: in particular, U.S.\ private fixed investment in information processing equipment (FRED B935RC1Q027SBEA) starts in 1995Q1, so the IT-GDP coefficient in the Dot-com Boom era is computed on $N=23$ paired log-returns rather than on the full $N=40$ era length. Table~\ref{tab:correlations} therefore reports the pair-specific effective $N$ alongside each coefficient; the era-window length $N$ in Section~\ref{sec:meth} should be read as an upper bound.

\paragraph{Operational definition of Strategic Decoupling.} We define ``Strategic Decoupling'' between two non-overlapping eras as a statistically significant change in the contemporaneous Pearson correlation between GDP growth and a technological proxy across the two eras, tested on the Fisher $r$-to-$z$ transform. Given sub-sample correlations $r_1$ (era 1, $N_1$ observations) and $r_2$ (era 2, $N_2$), with $z_i = \mathrm{atanh}(r_i)$ and joint standard error $\sigma = \sqrt{1/(N_1 - 3) + 1/(N_2 - 3)}$, the test statistic is $Z = (z_1 - z_2) / \sigma$ with two-sided $p$-value $2(1 - \Phi(|Z|))$. We apply this test between Cloud Expansion (era 1) and AI Era (era 2), with a sign restriction that the correlation must move from positive to negative (or vice versa) to qualify substantively as ``decoupling'' rather than mere attenuation.

\subsection{Cross-Correlation Function and Granger Causality}
To determine whether energy and data assets act as leading, coincident, or lagging indicators, we compute the Cross-Correlation Function (CCF) between GDP growth $(Y)$ and a technological proxy $(X)$ at lag $k$ as
\begin{equation}
r_{XY}(k) = \frac{\sum_t (X_{t-k} - \bar{X})(Y_t - \bar{Y})}{\sqrt{\sum_t (X_{t-k} - \bar{X})^2 \sum_t (Y_t - \bar{Y})^2}},
\end{equation}
where $k > 0$ indicates that $X$ leads $Y$ (predictive direction, $X_{t-k}$ enters as a forward-looking signal of $Y_t$), $k < 0$ indicates that $Y$ leads $X$ (reactive direction), and $k = 0$ is synchronous. We evaluate $k$ over $\{-4, -3, \ldots, +3, +4\}$ quarters. We use this sign convention consistently throughout Table~\ref{tab:lags} and Supplementary Table~S1; readers should note that the opposite sign convention is also common in the time-series literature.

We complement the CCF with Granger causality tests (\texttt{ssr\_ftest}) over a maximum lag of four quarters. We restrict the maximum lag from the eight quarters used in earlier drafts for two reasons: first, a year is the relevant horizon over which compute capacity might plausibly lead GDP given typical data-center build timelines; second, with only $N \approx 20$ post-COVID quarterly observations, a maximum lag of eight leaves too few residual degrees of freedom for the auxiliary regressions. Because we select the best lag from the set $\{1, 2, 3, 4\}$, we report both the minimum lag-wise $p$-value and a Bonferroni-adjusted $p$-value $p_{\text{Bonf}} = \min(1, 4 \cdot \min_k p_k)$. The headline Granger test (Supplementary Table~S1) is bidirectional between Irish data-center electricity consumption and Irish GDP, since the CCF profile in Table~\ref{tab:lags} is mixed about the leading direction.

\section{Data and Sources}

The dataset encompasses quarterly observations from major global economies, with a specialized focus on the Republic of Ireland as a proxy for hyper-digitized infrastructure.

\subsection{Variable Definitions}
We work with two parallel sets of series, paired with the GDP aggregate of the economy in which they are observed.
\begin{itemize}
    \item \textbf{U.S.\ GDP (GDP\_US):} Real Gross Domestic Product, chain-weighted, seasonally adjusted; FRED series GDPC1. This is the macroeconomic counterpart for all U.S.\ variables (IT investment, electricity PPI, semiconductor production and import series, and the U.S.\ renewables-generation proxy). Of these, three are reported as headline series in Table~\ref{tab:correlations}; semiconductor imports and the renewables proxy enter the cross-lag, Granger and spectral panels but not the headline contemporaneous correlations.
    \item \textbf{Irish GDP (GDP\_IE):} Real Gross Domestic Product at Constant Market Prices, seasonally adjusted; Central Statistics Office of Ireland matrix NAQ03, statistic NAQ03S04. This is the macroeconomic counterpart for the Irish data-center electricity series reported in Table~\ref{tab:lags}. We discuss in Section~\ref{sec:limitations} the well-known limitations of Irish headline GDP (Modified Gross National Income, GNI*, is the preferred domestic-activity aggregate) and re-running the same analysis on GNI* is on our agenda for confirmatory work.
    \item \textbf{IT Investment (IT\_US):} U.S.\ private fixed investment in information processing equipment, FRED series B935RC1Q027SBEA.
    \item \textbf{Electricity PPI (Elec\_US):} U.S.\ Producer Price Index for industrial electric power (FRED: WPS054321). This is a \emph{price} index; it is not the physical quantity of electricity consumed and is therefore not directly comparable to the metered consumption series used for Ireland (DC\_IE).
    \item \textbf{Semiconductor Metrics (Semi\_US):} Two U.S.\ series, the import price index for semiconductors (FRED: IR21320, ``Semi\_Import'') and a production/capacity proxy (FRED: CAPG3344S, ``Semi\_Production''), used as alternative measures of digital-inventory accumulation.
    \item \textbf{Renewables-generation proxy (PPA\_Proxy\_US):} FRED series IPN221114T8S, industrial production index for electric power generation from renewable sources (utility-scale, U.S., monthly, resampled to quarterly means). We use this as a coarse \emph{auxiliary} proxy for U.S.\ Power Purchase Agreement (PPA) capacity additions, available only from 2000Q1. It enters the cross-lag, Granger and spectral panels but is not part of the headline contemporaneous correlations in Table~\ref{tab:correlations}; the proxy is acknowledged to be an imperfect substitute for actual contracted PPA volumes, which are not published as a national quarterly series.
    \item \textbf{Data Center Load (DC\_IE):} Metered electricity consumption (gigawatt-hours per quarter) for the ``Data centres'' sector in Ireland, CSO matrix MEC02. This is a \emph{physical consumption} series.
\end{itemize}
The asymmetry between ``Elec\_US'' (a price index) and ``DC\_IE'' (a physical consumption series) is deliberate: the U.S.\ side lacks a publicly available quarterly metered series at comparable sectoral resolution, while CSO Ireland publishes data-center metered consumption directly. We flag the asymmetry explicitly so that Tables~\ref{tab:correlations} and~\ref{tab:lags} are not read as variations of a single ``electricity vs.\ GDP'' relationship.

\section{Results}
\label{sec:results}

This section presents the empirical results. We follow the IMRaD order: descriptive statistics and headline contemporaneous correlations across technological eras (Section~\ref{sec:results-eras}); the cross-lag analysis for the Irish case study (Section~\ref{sec:results-crosslag}); a substantive interpretation of the leads and lags (Section~\ref{sec:results-interp}); and Granger causality tests (Section~\ref{sec:results-granger}).

\subsection{Era-by-Era Contemporaneous Correlations}
\label{sec:results-eras}
Table~\ref{tab:correlations} reports synchronous Pearson correlation coefficients between U.S.\ real GDP growth and three U.S.\ counterpart growth series across selected technological eras.

\begin{table}
\caption{Synchronous Pearson correlation coefficients ($r$) with U.S.\ GDP Growth (Selected Eras), with raw two-sided $p$-values and Benjamini-Hochberg adjusted $p$-values $p_{\text{BH}}$ (pooled across all $\sim 120$ correlation tests reported in this paper). All three counterpart series are U.S.\ time series at quarterly frequency; ``Electricity (PPI)'' is the producer price index for industrial electric power (FRED: WPS054321), \emph{not} a physical consumption series. Reading: a cell entry of the form ``$r$ ; $p$ ($p_{\text{BH}}$)''. None of the reported tests reach the 5\% threshold after BH adjustment. \label{tab:correlations}}
\begin{center}
\resizebox{\textwidth}{!}{%
\begin{tabular}{lcccccc}
 & \multicolumn{2}{c}{\textbf{IT Investment}} & \multicolumn{2}{c}{\textbf{Electricity (PPI)}} & \multicolumn{2}{c}{\textbf{Semi Production}} \\
\textbf{Era} & $r$ & $p$ ($p_{\text{BH}}$) & $r$ & $p$ ($p_{\text{BH}}$) & $r$ & $p$ ($p_{\text{BH}}$) \\\hline
Dot-com Boom (1991Q2-2001Q1, $N=23$) & $+0.18$ & $0.40$ ($0.86$) & $-0.34$ & $0.12$ ($0.76$) & $-0.13$ & $0.55$ ($0.86$) \\
Web 2.0 Era (2001Q4-2007Q4, $N=24$) & $+0.26$ & $0.22$ ($0.82$) & $-0.09$ & $0.68$ ($0.94$) & $-0.02$ & $0.91$ ($0.96$) \\
Cloud Expansion (2009Q2-2019Q4, $N=42$) & $+0.33$ & $0.03$ ($0.72$) & $-0.44$ & $0.003$ ($0.28$) & $-0.26$ & $0.10$ ($0.76$) \\
\textbf{AI Era (2021Q1-2025Q4, $N=20$)} & $\mathbf{-0.25}$ & $0.29$ ($0.84$) & $\mathbf{+0.16}$ & $0.51$ ($0.86$) & $\mathbf{-0.42}$ & $0.07$ ($0.76$) \\
\end{tabular}%
}
\end{center}
\end{table}

The AI-Era point estimates for the United States show US IT investment and Semi production turning negative against US GDP ($r = -0.25$ and $r = -0.42$ respectively). At $N=20$, these coefficients fall well within the sampling distribution of zero correlation (two-sided 5\% critical value $\approx |r|=0.44$ for $df=18$), so we treat them as suggestive rather than statistically significant in isolation. The directional shift they indicate is, however, formally evaluated in Sections~\ref{sec:results-breakpoints}-\ref{sec:results-decoupling}: a Quandt-Andrews sup-Wald test locates a significant unknown break in the U.S.\ IT-GDP correlation at 2014Q3 (Table~\ref{tab:breakpoints}), and the corresponding Fisher $r$-to-$z$ test on pre/post-break sub-samples survives Bonferroni adjustment (panel (b) of Table~\ref{tab:decoupling}); by the operational definition in Section~\ref{sec:meth}, Strategic Decoupling is established \emph{at the empirical break date}, but \emph{not} at the imposed Cloud-vs-AI boundary. Two further clarifications affect how Table~\ref{tab:correlations} should be read. First, the ``Electricity'' column is the U.S. producer price index for industrial electricity (FRED: WPS054321), so it measures the \emph{price} of energy faced by U.S. firms, not their physical consumption; it is therefore not directly comparable to the metered consumption series used in Table~\ref{tab:lags}. Second, when Irish data-center electricity consumption is paired with Irish real GDP (CSO Ireland NAQ03S04, Constant Market Prices, SA) rather than US GDP, the synchronous correlation in the 2021Q1-2024Q4 window is large ($r = +0.41$, $N=16$) but, given a two-sided 5\% critical value of $|r| \approx 0.50$ at $df=14$, still does not reach conventional significance. These descriptive results motivate, but do not establish, a recomposition of the Kitchin cycle toward energy- and compute-backed assets.

\subsection{Cross-Lag Performance: Ireland as a Case Study Confounded by Multinational Footprint}
\label{sec:results-crosslag}
The dataset for the Republic of Ireland (2015Q1-2024Q4) provides the most granular view available of the energy-data nexus, although it now spans only 40 quarterly observations and ends one year earlier than the headline Irish national-accounts series. Throughout this subsection, Irish data-center metered electricity consumption (CSO matrix MEC02, sector ``Data centres'') is paired with Irish real GDP at Constant Market Prices, Seasonally Adjusted (CSO matrix NAQ03, statistic NAQ03S04). An earlier version of this analysis used U.S. real GDP as the macroeconomic counterpart; we discontinue that pairing because it conflates Ireland-hosted compute capacity with the U.S.\ macroeconomy and does not test the within-country lead-lag relationship implied by the ``New Energy and Data Cycle'' hypothesis.

\begin{table}
\caption{Cross-Correlation Coefficients for Irish GDP vs. Irish Data Center Electricity Consumption (log-returns, full sample 2015Q1-2024Q4). $r$ is the Pearson coefficient; $p$ the two-sided uncorrected $p$-value; $p_{\text{BH}}$ the Benjamini-Hochberg adjusted $p$-value, pooled across all 120 correlation tests reported in this paper; 95\% CI is the Fisher $z$-transform interval; $N$ is the number of paired log-return observations. $N$ varies across lags because shifting the data-center series by $|k|$ quarters drops $|k|$ overlapping observations. \label{tab:lags}}
\begin{center}
\begin{tabular}{lrrrcr}
Lag ($k$) & $r$ & $p$ & $p_{\text{BH}}$ & 95\% CI & $N$ \\ \hline
$k = -4$ (GDP leads DC by 4Q) & $+0.282$ & 0.101 & 0.760 & $[-0.06, +0.56]$ & 35 \\
$k = -3$ (GDP leads DC by 3Q) & $+0.231$ & 0.176 & 0.813 & $[-0.11, +0.52]$ & 36 \\
$k = -2$ (GDP leads DC by 2Q) & $+0.129$ & 0.447 & 0.863 & $[-0.20, +0.44]$ & 37 \\
$k = -1$ (GDP leads DC by 1Q) & $+0.115$ & 0.494 & 0.863 & $[-0.21, +0.42]$ & 38 \\
$k = \phantom{+}0$ (synchronous) & $+0.097$ & 0.555 & 0.863 & $[-0.22, +0.40]$ & 39 \\
$k = +1$ (DC leads GDP by 1Q) & $+0.251$ & 0.129 & 0.772 & $[-0.08, +0.53]$ & 38 \\
$k = +2$ (DC leads GDP by 2Q) & $-0.085$ & 0.618 & 0.916 & $[-0.40, +0.25]$ & 37 \\
$k = +3$ (DC leads GDP by 3Q) & $-0.218$ & 0.202 & 0.813 & $[-0.51, +0.12]$ & 36 \\
$k = +4$ (DC leads GDP by 4Q) & $-0.193$ & 0.267 & 0.838 & $[-0.49, +0.15]$ & 35 \\
\end{tabular}
\end{center}
\end{table}

No lag in Table~\ref{tab:lags} reaches a conventional 5\% significance level after multiplicity adjustment ($p_{\text{BH}} > 0.05$ for all nine entries); the smallest uncorrected $p$-value is $0.10$ at $k=-4$, but pooled across all 120 correlation tests run in the paper the Benjamini-Hochberg adjusted value rises to $0.76$. The corresponding two-sided 5\% critical value of the raw Pearson coefficient is $|r| \approx 0.33$ at $N \approx 36$. Substantively, none of the cross-correlation point estimates can be distinguished from zero at the sample size available.

\subsection{Interpretation of Leads and Lags}
\label{sec:results-interp}
With the small-sample caveat in mind, Table~\ref{tab:lags} does \emph{not} support a clean ``data center load leads GDP'' reading. Three observations follow:
\begin{enumerate}
    \item \textbf{Cross-correlation profile is asymmetric in the opposite direction to the original hypothesis.} The four largest cross-correlation coefficients in absolute value occur at $k=-4$ ($r=+0.28$), $k=+1$ ($r=+0.25$), $k=-3$ ($r=+0.23$), and $k=+3$ ($r=-0.22$), all values taken directly from Table~\ref{tab:lags}. The positive coefficients cluster on the $k<0$ side (GDP precedes data-center load), while the $k>+1$ side turns negative. This is more consistent with a \emph{reactive} or \emph{lagging} role for Ireland-hosted compute capacity, in which hyperscalers expand metered consumption a few quarters after macroeconomic expansion is observed, than with a leading-indicator interpretation.
    \item \textbf{Synchronous correlation in the AI Era is large but small-sample.} Restricting attention to the 2021Q1-2024Q4 sub-sample ($N=16$ after log-differencing) the synchronous correlation rises to $r = +0.41$. Given the two-sided 5\% critical value of $|r| \approx 0.50$ at $df=14$, this point estimate does not reach conventional significance but is the strongest single signal in the Irish data.
    \item \textbf{U.S.\ semiconductor patterns are unchanged by the re-pairing.} Independently of the Irish re-pairing, the contemporaneous correlation between U.S.\ semiconductor production and U.S.\ GDP turns negative in the AI Era ($r = -0.42$ in Table~\ref{tab:correlations}), consistent with semiconductor manufacturing increasingly following a longer-term Juglar-like cycle rather than the short Kitchin rhythm, although confirmation requires longer samples and frequency-domain analysis (e.g., spectral decomposition).
\end{enumerate}

\paragraph{HAC inference on the cross-lag profile.} The 95\% confidence intervals in Table~\ref{tab:lags} use the Fisher $z$-transform under an i.i.d.\ bivariate-normal assumption. Quarterly macroeconomic log-returns can retain mild residual autocorrelation even after differencing, so we recompute the standard errors with a Newey-West HAC adjustment (Bartlett kernel on $u_t = x^{*}_t y^{*}_t$ with $x^{*}, y^{*}$ sample-standardised, lag $L = \lfloor 4 (N/100)^{2/9} \rfloor$, which yields $L = 3$ on the available Irish sample). Full output is in \verb|hac_results.csv|. The largest ``GDP precedes data-centre load'' coefficient at $k = -3$ ($r = +0.23$) becomes statistically significant under HAC inference ($p_{\text{HAC}} = 0.009$ uncorrected, raw Bonferroni across the nine lags $\approx 0.08$), as does the mirror coefficient at $k = +3$ ($r = -0.22$, $p_{\text{HAC}} = 0.041$ uncorrected, Bonferroni $\approx 0.37$); for the synchronous lag the HAC SE is similar to the i.i.d.\ SE, and for the $k = -4$ peak emphasised above the HAC SE leaves the picture essentially unchanged ($p_{\text{HAC}} = 0.084$ uncorrected). The qualitative reading is preserved: point-estimate mass remains on the $k < 0$ side (GDP precedes data-centre load), and HAC inference, if anything, sharpens this asymmetry while leaving no single lag significant after multiplicity correction at the available sample size.

\subsection{Granger causality}
\label{sec:results-granger}
Bidirectional Granger causality tests between Irish data-centre electricity and Irish GDP with maximum lag $K=4$ quarters (full table reported as Table~S1 in the Supplementary Materials) do not pass a conventional 5\% threshold after Bonferroni adjustment over the four lag choices: the DC-on-GDP direction has the smallest uncorrected $p$-value of the headline tests ($p=0.077$ at $k=1$, $p_{\text{Bonf}} = 0.31$), the reverse direction is weaker ($p=0.14$ at $k=4$), and AI-Era-only restriction ($N=16$) eliminates the modest signal entirely. The picture is consistent with the cross-correlation Table~\ref{tab:lags}: the available sample is too short to disentangle directionality, and we do not interpret these results as supporting a forecasting role for data-centre load.

\subsection{Structural Breakpoint Tests}
\label{sec:results-breakpoints}
The era boundaries used in Sections~\ref{sec:results-eras}-\ref{sec:results-crosslag} are taken from technology history and are imposed rather than estimated. To check whether they correspond to genuine empirical breaks, we run Quandt-Andrews sup-Wald tests for an unknown breakpoint in the contemporaneous correlation between each headline pair, with trimming proportion $\pi_0 = 0.15$. \citet{andrews1993} reports an asymptotic 5\% critical value of $\approx 8.85$ for one parameter and 15\% trim, which we use as the rejection threshold; a parametric bootstrap $p$-value would tighten this assessment but is not implemented in the present specification.

\begin{table}
\caption{Quandt-Andrews sup-Wald Tests for an Unknown Breakpoint in Contemporaneous Correlation. ``Break date'' is the maximiser of the Wald statistic; $r_{\text{pre}}$ and $r_{\text{post}}$ are the sub-sample correlations. Threshold for rejection of constant correlation at 5\% is sup-W $\approx 8.85$ (Andrews 1993, $q=1$, $\pi_0 = 0.15$). \label{tab:breakpoints}}
\begin{center}
\begin{tabular}{lrrrrr}
Pair & $N$ & sup-W & Break date & $r_{\text{pre}}$ & $r_{\text{post}}$ \\ \hline
US GDP vs US IT investment & 122 & 9.69 & 2014-07-01 & $+0.45$ & $-0.12$ \\
US GDP vs US semiconductor production & 122 & 12.66 & 2020-07-01 & $+0.28$ & $-0.54$ \\
IE GDP vs IE data-center electricity & 39 & 2.68 & 2021-01-01 & $-0.14$ & $+0.41$ \\
\end{tabular}
\end{center}
\end{table}

Three findings emerge. First, the US IT-GDP correlation shows a statistically significant break in mid-2014 (sup-W $= 9.69$, above the 5\% threshold), with the correlation falling from $+0.45$ pre-break to $-0.12$ post-break. The break date is several years \emph{before} the start of our nominal AI Era window (2021Q1), implying that the ``Strategic Decoupling'' of IT investment from US GDP is a longer-running phenomenon than the AI-era reading would suggest, and one that is at most partly attributable to the AI investment cycle. Second, the US semiconductor production-GDP correlation has the strongest break of the three pairs (sup-W $= 12.66$), dated to mid-2020, which is well above the 5\% threshold and aligns naturally with COVID-era supply shocks and the early stages of the post-pandemic ``silicon cycle.'' Third, the Irish data-center electricity-GDP correlation has its sup-W maximiser at 2021Q1, which coincides exactly with our nominal AI Era boundary, but the test statistic itself (2.68) falls well below the critical value: with only 39 paired observations, we cannot reject the hypothesis of a constant Irish correlation throughout 2015-2024. We therefore continue to treat the Irish results as exploratory; the boundary at 2021Q1 is, on the available data, a plausible but unverified marker rather than an empirically established break.

\subsection{Strategic Decoupling Tests}
\label{sec:results-decoupling}
The structural breakpoint tests above estimate where, on the time axis, the contemporaneous correlation between two series shifts. A complementary question is whether the correlation observed before a candidate boundary differs significantly from that observed after. Following the operational definition introduced in Section~\ref{sec:meth}, we apply a Fisher $r$-to-$z$ test to each headline pair under two boundary specifications (Table~\ref{tab:decoupling}): in panel (a), the \emph{imposed} Cloud Expansion vs AI Era boundary; in panel (b), the \emph{empirical} Quandt-Andrews break date for each pair (read directly from Table~\ref{tab:breakpoints}). Reading both panels side by side isolates the effect of imposing era boundaries that do not coincide with the data-identified regime change.

\begin{table}
\caption{Strategic Decoupling Tests (Fisher $r$-to-$z$). Panel (a) uses the \emph{imposed} Cloud Expansion (2009Q2-2019Q4) vs AI Era (2021Q1-2025Q4) boundary; panel (b) uses the \emph{empirical} Quandt-Andrews break date for each pair (Table~\ref{tab:breakpoints}). $z$ is the standardised statistic; $p$ the two-sided uncorrected $p$-value; $p_{\text{Bonf}}$ is the Bonferroni-adjusted $p$-value computed \emph{within each panel} (four tests in panel (a), three tests in panel (b)). \label{tab:decoupling}}
\begin{center}
\resizebox{\textwidth}{!}{%
\begin{tabular}{lrrrrrrrr}
Pair & $r_{\text{pre}}$ & $N$ & $r_{\text{post}}$ & $N$ & $\Delta r$ & $z$ & $p$ & $p_{\text{Bonf}}$ \\ \hline
\multicolumn{9}{l}{\textit{(a) Imposed boundary: Cloud Expansion vs AI Era}} \\
US GDP vs US IT investment   & $+0.329$ &  42 & $-0.247$ & 20 & $-0.576$ & $+2.04$ & $0.041$ & $0.164$ \\
US GDP vs US electricity PPI & $-0.441$ &  42 & $+0.155$ & 20 & $+0.597$ & $-2.17$ & $0.030$ & $0.120$ \\
US GDP vs US semi production & $-0.257$ &  42 & $-0.419$ & 20 & $-0.162$ & $+0.63$ & $0.528$ & $1.000$ \\
IE GDP vs IE DC electricity  & $-0.129$ &  19 & $+0.415$ & 16 & $+0.543$ & $-1.53$ & $0.127$ & $0.506$ \\ \hline
\multicolumn{9}{l}{\textit{(b) Empirical boundary: Quandt-Andrews break date per pair}} \\
US IT-GDP, pre/post 2014Q3   & $+0.448$ &  76 & $-0.116$ & 46 & $-0.564$ & $+3.11$ & $0.002$ & $\mathbf{0.006}$ \\
US semi-GDP, pre/post 2020Q3 & $+0.281$ & 100 & $-0.539$ & 22 & $-0.821$ & $+3.56$ & $0.0004$ & $\mathbf{0.001}$ \\
IE DC-GDP, pre/post 2021Q1   & $-0.141$ &  23 & $+0.415$ & 16 & $+0.556$ & $-1.64$ & $0.102$ & $0.305$ \\
\end{tabular}%
}
\end{center}
\end{table}

In panel (a), two of the four imposed-boundary pairs show a difference in correlation between Cloud Expansion and AI Era that reaches an uncorrected 5\% significance level: US IT investment ($p = 0.041$) and US electricity PPI ($p = 0.030$). The signs are economically sensible: the IT-GDP correlation shifts from positive ($+0.33$) to negative ($-0.25$), which matches the substantive ``decoupling'' definition; and the electricity-PPI-GDP correlation shifts from clearly negative ($-0.44$) to mildly positive ($+0.16$), which is consistent with energy prices and output behaving like joint outputs of post-pandemic supply shocks rather than the textbook inverse relationship of the Cloud era. Both $p$-values would, however, fall above the conventional 5\% threshold after multiplicity correction across the four pairs ($p_{\text{Bonf}} = 0.164$ for IT and $p_{\text{Bonf}} = 0.120$ for electricity PPI). The Irish DC-GDP pair shows the largest substantive change in panel (a) ($\Delta r = +0.54$), but at only $N_1 = 19$ and $N_2 = 16$ the test is under-powered ($p = 0.127$). The US semiconductor production-GDP pair shows essentially no era-level change ($\Delta r = -0.16$, $p = 0.53$) in panel (a), even though Table~\ref{tab:breakpoints} identified a significant break in mid-2020; the apparent contradiction is consistent with the break occurring \emph{within} Cloud Expansion rather than at the Cloud-vs-AI boundary, and serves as a reminder that imposed era boundaries can mask, rather than reveal, genuine structural change.

Panel (b) confirms this reading directly. Replacing the imposed Cloud-vs-AI boundary with the empirically estimated Quandt-Andrews break date sharpens both U.S.\ tests substantially. The U.S.\ IT-GDP correlation pre/post 2014Q3 ($r_{\text{pre}} = +0.448$, $N = 76$; $r_{\text{post}} = -0.116$, $N = 46$) yields a Fisher $z = +3.11$ ($p = 0.002$), which \emph{survives} Bonferroni adjustment within the empirical-break family of three tests ($p_{\text{Bonf}} = 0.006$). The U.S.\ semi-GDP correlation pre/post 2020Q3 yields an even stronger $z = +3.56$ ($p = 0.0004$, $p_{\text{Bonf}} = 0.001$). The Irish DC-GDP pre/post 2021Q1 test, by contrast, remains non-significant ($z = -1.64$, $p = 0.102$), consistent with the sup-W $= 2.68$ reported in Table~\ref{tab:breakpoints} being well below the Andrews critical value of $\approx 8.85$. The substantive implications are twofold: (i) by the operational definition of Section~\ref{sec:meth}, ``Strategic Decoupling'' is statistically established for U.S.\ IT-GDP and U.S.\ semi-GDP \emph{at their data-identified break dates}, but not at the imposed Cloud-vs-AI boundary; and (ii) the U.S.\ IT-GDP regime change predates the nominal AI Era by approximately six years, so the AI Era is best read as the continuation of a longer-running decoupling rather than as its onset. We caution that the empirical break dates are themselves selected from the data, so $p$-values in panel (b) should be interpreted in the spirit of the sup-Wald test (already significant for the U.S.\ pairs in Table~\ref{tab:breakpoints}) rather than as additional independent evidence.

\subsection{Robustness: Choice of Irish Activity Measure}
\label{sec:results-robustness}
The headline AI-Era result ``Irish data-center electricity correlates synchronously with Irish GDP at $r = +0.41$'' depends on which Irish activity aggregate we use as the right-hand-side variable. Irish real GDP is well known to be inflated by intra-firm activity of large multinationals (the so-called ``leprechaun economics'' problem); real GNP, by contrast, deducts net factor income remitted to the rest of the world and is closer to a domestic-activity measure. We therefore re-run the synchronous AI-Era correlation against three alternative activity aggregates available in CSO matrix NAQ03 (Table~\ref{tab:robustness}).

\begin{table}
\caption{Robustness of the synchronous AI-Era correlation between Irish data-center electricity consumption and three Irish activity measures (Real GDP, Real GNP, Real GVA from CSO matrix NAQ03), with a 5\% winsorisation variant and the full-sample 2015-2024 comparison. A separate ``without COVID 2020 quarters'' variant is not reported because the AI Era window already starts in 2021Q1 and therefore excludes 2020. All correlations are computed on log-returns. \label{tab:robustness}}
\begin{center}
\resizebox{\textwidth}{!}{%
\begin{tabular}{llrrcr}
Activity measure & Sample & $r$ & $p$ & 95\% CI & $N$ \\ \hline
Real GDP (NAQ03S04) & AI Era                & $+0.415$ & 0.110 & $[-0.10, +0.76]$ & 16 \\
Real GDP (NAQ03S04) & AI Era + 5\% winsor   & $+0.427$ & 0.099 & $[-0.09, +0.76]$ & 16 \\
Real GDP (NAQ03S04) & Full sample 2015-2024 & $+0.097$ & 0.555 & $[-0.23, +0.40]$ & 39 \\
\textbf{Real GNP (NAQ03S06)} & \textbf{AI Era}     & $\mathbf{+0.008}$ & 0.977 & $[-0.49, +0.50]$ & 16 \\
Real GNP (NAQ03S06) & AI Era + 5\% winsor   & $+0.042$ & 0.877 & $[-0.46, +0.53]$ & 16 \\
Real GNP (NAQ03S06) & Full sample 2015-2024 & $-0.103$ & 0.531 & $[-0.41, +0.22]$ & 39 \\
Real GVA (NAQ03S01) & AI Era                & $+0.434$ & 0.093 & $[-0.08, +0.76]$ & 16 \\
Real GVA (NAQ03S01) & AI Era + 5\% winsor   & $+0.447$ & 0.083 & $[-0.06, +0.77]$ & 16 \\
Real GVA (NAQ03S01) & Full sample 2015-2024 & $+0.102$ & 0.539 & $[-0.22, +0.40]$ & 39 \\
\end{tabular}%
}
\end{center}
\end{table}

The substantive finding is that the headline correlation is not robust to the choice of activity aggregate. Pairing data-center electricity with Real GDP gives $r = +0.41$; pairing the same series with Real GVA gives a very similar $r = +0.43$; but pairing it with Real GNP \emph{collapses the correlation to $r \approx 0$}. Since GNP differs from GDP and GVA mainly by deducting net factor income paid to non-residents, which in Ireland is dominated by repatriated profits of large multinationals headquartered abroad, the natural interpretation is that the AI-Era correlation between Irish data-center load and Irish GDP/GVA is driven by the multinational footprint shared by both series (hyperscaler operations are simultaneously a contributor to Irish gross output and a major consumer of Irish electricity), rather than by a cyclical relationship between data-center capacity and domestically generated economic activity. The 5\% winsorisation has only a marginal effect (the AI-Era window contains too few extreme observations for trimming to bite), and the COVID-restriction column is identical to the AI-Era column because the AI Era starts in 2021Q1 and therefore already excludes the 2020 quarters.

We treat this as a strong qualifier on Section~\ref{sec:results-crosslag}: the appropriate confirmatory replication should use GNI* (CSO Ireland's officially preferred ``modified'' aggregate, available only at annual frequency) or a panel of data-center hubs less dependent on multinational tax structures, and not headline Irish GDP.

\subsection{Robustness: Seasonal Adjustment of DC\_IE Load}
\label{sec:results-stl}

A STL-based seasonal-adjustment robustness check on the headline 9-lag DC\_IE-GDP cross-correlation profile of Table~\ref{tab:lags} (full table reported as Table~S2 in the Supplementary Materials; STL period $=4$, robust mode, $N = 40$) preserves the qualitative reading of Section~\ref{sec:results-crosslag}. Specifically, seasonal adjustment \emph{strengthens} the largest ``GDP precedes data-centre load'' coefficient at $k = -4$ ($r_{\text{SA}} = +0.39$, $p = 0.021$ uncorrected, versus $r_{\text{NSA}} = +0.28$, $p = 0.101$) and uniformly moves the ``DC leads GDP'' side ($k > 0$) toward zero, consistent with the original ``DC leads'' coefficients being partly an aliasing artefact of the annual seasonal in DC\_IE rather than evidence of a forecasting role for data-centre load. The AI-Era synchronous correlation drops modestly from $r = +0.41$ (NSA) to $r = +0.35$ (SA, $N = 16$), and the GNP-collapse documented in Section~\ref{sec:results-robustness} is unaffected by seasonal adjustment because GNP-driven multinational tax-structure dynamics operate on the trend component, not on the annual seasonal.

\subsection{Frequency-Domain Evidence: Where Is the Kitchin Cycle?}
\label{sec:results-spectral}

\begin{figure}[ht]
\begin{center}
\includegraphics[width=0.80\textwidth]{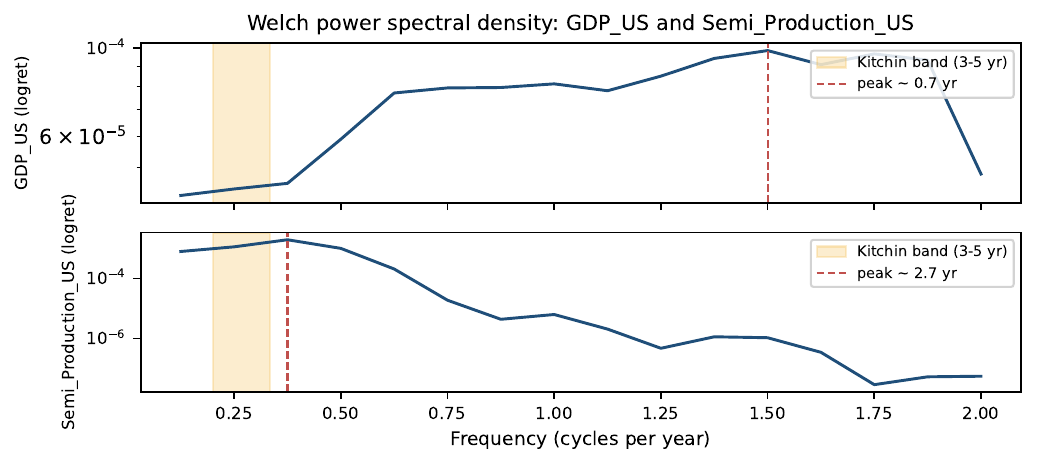}
\end{center}
\caption{Welch power spectral density for the two headline series most relevant to the Kitchin question: U.S.\ real GDP and U.S.\ semiconductor production (log-returns). Frequency is in cycles per year; PSD on $\log_{10}$ axis. The shaded band marks the canonical 3-5 year Kitchin horizon; the dashed line marks the dominant (\#1) spectral peak in each series. The full six-panel PSD for all headline log-return series is reported as Figure~S2 in the Supplementary Materials and as \texttt{spectral\_results.csv}; numeric peaks for every series are reported in Table~\ref{tab:spectral}. Sample window: 1995Q1-2025Q4 ($N \approx 123$). \label{fig:spectral}}
\end{figure}

The era-by-era and cross-correlation evidence above describes \emph{contemporaneous} comovement, but the Kitchin-cycle hypothesis is a claim about the dominant \emph{periodicity} of the underlying series. As a complementary test, we compute the Welch power spectral density of each headline log-return series, with sampling frequency $f_s = 4$ cycles per year (quarterly data), Hann window, segment length $n_{\text{perseg}} = \min(32, N)$ and 50\% overlap. The canonical Kitchin range of 3-5 years corresponds to a frequency band of $0.20$ to $0.33$ cycles per year, which we mark on Figure~\ref{fig:spectral}. Table~\ref{tab:spectral} reports, for each series, the period (in years) of the three largest non-zero spectral peaks.

\begin{table}[h]
\caption{Top-3 spectral peaks of headline log-return series (Welch PSD, 1995Q1-2025Q4 for U.S.\ series, 2015Q1-2024Q4 for Irish DC). ``Period'' is the inverse of the peak frequency, expressed in years; the canonical Kitchin band is 3-5 years. \label{tab:spectral}}
\begin{center}
\resizebox{\textwidth}{!}{%
\begin{tabular}{lrrrr}
Series (log-returns) & Peak \#1 (yr) & Peak \#2 (yr) & Peak \#3 (yr) & In Kitchin band? \\ \hline
GDP\_US             & 0.67 & 0.57 & 0.73 & no \\
GDP\_IE             & 0.53 & 0.57 & 0.62 & no \\
IT\_US              & 1.60 & 2.00 & 0.57 & below band \\
Semi\_Production\_US & 2.67 & \textbf{4.00} & 2.00 & \textbf{Peak \#2 inside band} \\
Electricity\_PPI\_US & 0.80 & 1.33 & 0.73 & no \\
DC\_IE              & 1.14 & 1.33 & 1.00 & no (annual seasonality) \\
\end{tabular}%
}
\end{center}
\end{table}

Three observations follow:

\begin{enumerate}
\item \textbf{Aggregate output series do not show a Kitchin peak.} For both U.S.\ and Irish real GDP, the top three spectral peaks are concentrated at periods of 0.53-0.73 years (roughly 2-3 quarters), well to the right of the Kitchin band. On the present sample, the 3-5 year inventory-cycle frequency is not the dominant rhythm of quarterly GDP growth. This is consistent with the broader empirical literature that argues the classical Kitchin signature has faded in modern post-1980 macro data (see, e.g., the discussion in \citet{burns1946} of the original NBER reference-cycle methodology and subsequent inventory-cycle re-evaluations).

\item \textbf{Semiconductor production is the one series with a peak inside the Kitchin band.} U.S.\ semiconductor production has its second-largest spectral peak at exactly 4.0 years, the centre of the 3-5 year Kitchin window, and its largest peak at $\approx 2.7$ years (around the Juglar-Kitchin boundary). This is consistent with the ``silicon cycle'' discussion in Section~\ref{sec:discussion-silicon}: semiconductor capacity does retain a measurable medium-term cyclical signature, while the macroeconomic aggregates it co-moves with do not.

\item \textbf{Irish data-center electricity is dominated by annual seasonality.} The dominant peak in DC\_IE is at $\approx 1.14$ years, with secondary peaks at 1.33 and 1.00 years; all three are consistent with an annual cooling/heating cycle rather than with a multi-year inventory or investment cycle. Together with the GNP-collapse result in Section~\ref{sec:results-robustness}, this suggests that any AI-era signal in Irish data-center load is more naturally read as the sum of (a) an annual energy-demand seasonal and (b) a multinational tax-structure level shift, rather than as a recomposed Kitchin cycle proper.
\end{enumerate}

Taken together, the frequency-domain evidence \emph{tempers} the headline narrative: the post-2021 ``decoupling'' described in Sections~\ref{sec:results-eras}-\ref{sec:results-decoupling} is genuine in the time domain but does not yet correspond to a clean Kitchin-period peak in the underlying series; the one variable that does sit in the Kitchin band, U.S.\ semiconductor production, behaves as if it is migrating toward the Juglar window, in line with the substantive interpretation in Section~\ref{sec:discussion-silicon}.

\section{Discussion: Implications for Macroeconomic Measurement and Forecasting}
\label{sec:discussion}

If a ``Digital-Energy Kitchin Cycle'' is operative, it has implications for how macroeconomic forecasters and policy analysts should select and weight leading indicators in data-intensive economies. We outline the implications below as conjectures consistent with our exploratory findings; each requires confirmatory work before being treated as a basis for portfolio or policy decisions.

\subsection{Energy as a candidate \emph{coincident} indicator}
On Irish data, the leading-side cross-correlation at $k=+4$ between Irish data-center electricity consumption and Irish GDP is \emph{negative} ($r=-0.20$), while the largest positive coefficient is at $k=-4$, where GDP precedes data-center load. The defensible reading is that data-center metered consumption in Ireland is a candidate \emph{coincident or moderately lagging} indicator of macroeconomic activity, with the strongest signal in the synchronous AI Era window ($r=+0.41$, $N=16$). Utility-scale load forecasts and Power Purchase Agreement (PPA) volumes in technology hubs (e.g., Ireland, the U.S.\ Mid-Atlantic data-center cluster) remain plausible inputs to nowcasting frameworks such as that of \citet{aruoba2009}, but our results do not establish a forecasting role for these series, and any such claim should be tested directly on longer monthly samples with multiple-testing controls.

\subsection{Apparent decoupling of the silicon cycle}
\label{sec:discussion-silicon}
The negative contemporaneous correlation between semiconductor production and GDP in the AI Era ($r=-0.42$), reinforced by a significant Quandt-Andrews break in mid-2020 (Table~\ref{tab:breakpoints}) and by a spectral peak at $\approx 4$ years (Table~\ref{tab:spectral}), is consistent with the ``Silicon Cycle'' increasingly behaving as a strategic, longer-term capital-formation series rather than as a short-cycle inventory series. In the language of \citet{schumpeter1939}, the silicon cycle appears to be migrating from the Kitchin (short, inventory-driven) wave toward the Juglar (medium, capital-equipment-driven) wave. If confirmed in larger samples, this would imply that semiconductor capacity should be treated more in the spirit of \citet{dixit1994}-style real-options investment than as a Kitchin-cycle inventory adjustment.

\section{Limitations}
\label{sec:limitations}

The empirical results presented in this paper should be read as exploratory rather than confirmatory. Several limitations qualify their interpretation.

\paragraph{Sample size.} The post-2021 ``Strategic Decoupling'' window contains $N=20$ quarterly observations ($df=18$, two-sided 5\% critical $|r| \approx 0.44$). None of the headline AI-Era coefficients ($r=-0.25$ for IT-GDP, $r=-0.19$ for $k=+4$ in Table~\ref{tab:lags}, $r=-0.42$ for Semi production) reach this threshold individually, although several are jointly suggestive. Replication on longer monthly samples and additional economies is required before any pattern can be regarded as established.

\paragraph{Irish GDP measurement.} Irish GDP is well known to be heavily influenced by intra-firm activity of large multinationals \citep{fitzgerald2018, lane2017, cso2017}, including intellectual-property relocations that produced an officially reported real GDP growth of 26\% in 2015. The preferred domestic-activity aggregate, GNI*, is only available annually and cannot serve as a quarterly counterpart here. Section~\ref{sec:results-robustness} shows that pairing Irish data-center electricity with quarterly Real GNP (conceptually closer to GNI*) collapses the AI-Era headline correlation from $r = +0.41$ to $r \approx 0$. Confirmatory replication on GNI*, modified domestic demand, or a panel of data-center hubs less dominated by multinational tax structures remains the natural next step.

\paragraph{Era boundaries are imposed, not estimated.} Our six technological eras are taken from the descriptive technology-history literature. The Quandt-Andrews tests in Section~\ref{sec:results-breakpoints} identify empirically significant breaks at 2014Q3 (US IT-GDP) and 2020Q3 (US semi-GDP) but cannot reject constant correlation on the smaller Irish sample, and Section~\ref{sec:results-decoupling}, panel (b), shows that Fisher $r$-to-$z$ at these empirical breaks yields significant decoupling for both U.S.\ pairs after Bonferroni adjustment within the empirical-break family, while the panel (a) imposed-boundary tests do not. Cloud Expansion therefore conflates a pre-decoupling sub-regime (1995Q1-2014Q2) with an early post-decoupling sub-regime (2014Q3-2019Q4), and the headline Cloud-era $r = +0.33$ should be read as a weighted average rather than as a stable-regime estimate. A future Bai-Perron multiple-break analysis on the rolling correlations would offer a still more disciplined characterisation; panel (b) of Table~\ref{tab:decoupling} is reported in the spirit of confirming the Quandt-Andrews finding rather than as an additional independent test.

\paragraph{Inference assumptions.} The Fisher $z$-transform confidence intervals in Tables~\ref{tab:lags} and~\ref{tab:robustness} assume i.i.d.\ bivariate-normal observations; mild residual autocorrelation in quarterly log-returns makes them approximate. Section~\ref{sec:results-interp} reports Newey-West HAC standard errors for the headline Irish cross-lag profile (\verb|hac_results.csv|): the qualitative reading is preserved, and HAC inference sharpens the asymmetric ``GDP precedes data-centre load'' pattern. Extending HAC to Tables~\ref{tab:correlations} and~\ref{tab:robustness} is a natural confirmatory step left for future work.

\paragraph{Stationarity and spurious correlation.} ADF and KPSS tests on every series in log-levels and log-returns are reported in \verb|stationarity_tests.csv|. After log-differencing, ADF rejects the unit-root null at 5\% for every series, but the U.S.\ semiconductor import and production indices retain a KPSS rejection at $p = 0.01$ even on log-returns. Second-differencing the import index makes it unambiguously stationary (ADF $p < 0.001$, KPSS $p \approx 0.10$); the production index improves on KPSS (statistic $1.26 \to 0.04$) but its ADF $p$-value rises to $0.07$, so it is best described as ``weakly identified.'' Conclusions involving either semiconductor variable should be discounted accordingly, and a replication on physical-volume measures (wafer-starts, shipments) is the natural next step.

\paragraph{Causality.} Our cross-correlation results are descriptive. Bidirectional Granger tests between Irish data-center electricity and Irish GDP (Supplementary Table~S1, max lag $K=4$, Bonferroni over the lag set) do not pass 5\% after multiplicity adjustment, and the AI-Era-only restriction leaves the headline test under-powered. The current sample is too short to disentangle direction of causation; future work should use longer monthly series, panel methods across multiple data-center hubs, and IV strategies exploiting exogenous variation in electricity supply (e.g., the EirGrid Dublin grid-connection moratorium, PPA pricing shocks).

\paragraph{Confounding by COVID-19.} The 2021-2025 AI Era window overlaps with persistent pandemic-related disruptions: 2021Q1-Q3 (reopening volatility), 2021Q4 (Omicron), and 2022Q1 (Omicron tail plus Russia-Ukraine energy shock) are the most directly affected quarters. Disentangling pandemic effects from a structural ``AI arms race'' shift would require explicit pandemic dummies, counterfactual decomposition, or pre-/post-pandemic instrument strategies, none of which is attempted here.

\section{Conclusion}
\label{sec:conc}

This paper has revisited the Kitchin cycle in light of the digital and AI-driven transformation of the global economy. Drawing on quarterly U.S.\ (FRED) and Irish (CSO) data, we documented three patterns consistent with a recomposition of the cycle around energy and compute. First, the U.S.\ contemporaneous IT-GDP correlation shifts from positive in earlier eras ($r=+0.33$ during Cloud Expansion) to negative in the AI Era ($r=-0.25$, $N=20$), with a Quandt-Andrews break at 2014Q3 and a Fisher $r$-to-$z$ test on the empirical break that survives Bonferroni ($z = +3.11$, $p_{\text{Bonf}} = 0.006$); the analogous Cloud-vs-AI imposed-boundary test does not ($p_{\text{Bonf}} = 0.164$), so ``Strategic Decoupling'' is a longer-running, pre-AI phenomenon. Second, the Irish AI-Era synchronous correlation between data-centre electricity and Real GDP ($r=+0.41$, $N=16$) \emph{collapses to $r \approx 0$ when Real GDP is replaced by Real GNP}, indicating that the headline is largely an artefact of a shared multinational tax-structure footprint; the cross-correlation profile is also asymmetric on the ``GDP precedes data-centre load'' side ($k=-4$, $r=+0.28$, vs.\ $k=+4$, $r=-0.19$), consistent with compute capacity as a coincident or moderately lagging indicator. Third, U.S.\ semiconductor production shows a progressively negative GDP correlation across post-1991 eras ($r=-0.13 \to -0.42$), a Quandt-Andrews break at 2020Q3, and a spectral peak at 4 years inside the Kitchin band, while neither U.S.\ nor Irish GDP shows a comparable peak.

The evidence remains exploratory: only two of $\sim 120$ correlation tests cross an uncorrected 5\% threshold, none survive Benjamini-Hochberg FDR adjustment, and the era boundaries are partly empirical (U.S.) and partly imposed (Irish). Three confirmatory extensions are valuable: monthly-frequency replication across data-centre hubs (Ireland, the Netherlands, Singapore, Northern Virginia, Frankfurt); re-estimating Irish results against GNI* or a multinational-adjusted GDP series; and identification strategies exploiting exogenous variation in grid capacity and PPA pricing (e.g., the EirGrid Dublin connection moratorium). If these patterns survive, the Kitchin cycle of the next decade may be best monitored not by warehouse stock-to-sales ratios but by dashboards of contracted electricity and compute; if not, the more conservative conclusion is that the cycle has dematerialised faster than the data infrastructure required to measure it.

\section{Acknowledgments}
OpenAI ChatGPT [GPT 5.5, 04-11 V 2026] was used to support manuscript-level consistency checks and code-related work. Kimi by Moonshot AI [Kimi 2.6, 04-11 V 2026 ] was used as an independent validation tool for code and substantive review of the manuscript. DeepSeek [v4-pro, 04-11 V 2026 ] was used for mathematical, statistical, and substantive verification of the manuscript. Google NotebookLM [Grounded Synthesis Engine, May 2026] was used as the environment for source-grounded synthesis and cross-document reasoning. Cursor [default model April 2026] was used to develop the first version of our Python code, employing a "vibecoding" approach to rapid prototyping. The use of these tools was restricted to assistance with language clarity, coherence checking, code review, and independent validation. The tools were not used to autonomously conduct the research, determine the conclusions, generate synthetic data, or substitute for primary-source verification. All empirical results reported in the manuscript derive from the authors' stated data sources and reproducible analytical pipeline. All AI-assisted suggestions were treated as provisional and were accepted only after human review.

\bibliographystyle{Chicago}

\bibliography{FIN_AI}
\end{document}